# Geometrical formulation of quantum fields

S. R. Vatsya

648 Inverness Ave., London, Ontario, Canada, N6H 5R4
e-mail: raj.vatsya@gmail.com  Phone: (1) 519 474 1183

## Abstract

Path integral formulation of quantum mechanics defines the wavefunction associated with a particle as a sum of phase-factors, which are periodic functions of classical action. In the present article, this periodicity is shown to impart the corresponding periodicity to a one parameter family of wavefunctions generated by the translations of arclength used to parameterize the trajectories. Translation parameter is adjoined to the base to obtain an extended manifold. Periodicity of the family of wavefunctions with respect to the translation parameter together with solutions of the generalized Klein-Gordon equation, which is deducible from the path integral formulation, is used to define a quantized field with zero vacuum energy in the extended manifold with the particle being its quantum. Classical description of essentially the same particle is obtained in the extended higher dimensional space using its properties. This manifold can replace the base in this treatment to continue the program for higher dimensional manifolds generated in the process. Results are illustrated by taking the three-dimensional Euclidean space the base, which yields the classical and quantum particle descriptions of photon in the base and the field description in the resulting extended manifold, which is identified with the Minkowski spacetime. The field formulation yields the quantized Maxwell's equations. A novel interpretation of time as the corresponding translation parameter results in the process. Classical description in the extended manifold, i.e., the Minkowski spacetime, results in the relativistic description of a massive particle related to the photon. The results are further illustrated for this massive particle in the Minkowski spacetime obtaining parallel results.

## 1. Background

In this section, we collect some known results required to develop the material in following sections. Only the brief descriptions are given; details are available elsewhere.

### 1.1. Feynman's Path Integral Formulation

In Feynman' s the path integral formulation of quantum mechanics [1-3], the particle wavefunction $\psi$ is defined by [1, pp. 29, 57-58]

$$\psi[x,\ \tau(x)]\ =\ \sum_{\text{all paths}} \exp\{\ i\ S[\rho_{xx_0}(\tau)]\},\qquad (1)$$

where $S[\rho_{xx_0}(\tau)]$ is the classical action along a trajectory $\rho_{xx_0}(\tau)$ from $x_0$ to $x$ in the underlying $N$-dimensional manifold $\mathcal{R}_N$ taken to be a Riemannian space. The trajectories are parameterized by arclength $\tau$ treated as an independent parameter to accommodate a collection of trajectories [4-6]. The initial point varies over the set $\{x_0\}$ of the initial points of all trajectories with a fixed parameter value $\tau(x_0)$, and $\tau(x)$ is the parameter value at $x$. The sum in (1) assigns equal weight to each term, which is absorbed in it. The space in Feynman's original formulation was taken to be 3D Euclidean; generalized form (1) given here was obtained later.

An integral representation of the wavefunction

$$\psi(x,\bar{\tau}+\varepsilon)\ =\ \int d\tilde{m}(y)\ \exp\left[i\ \hat{S}(x,\bar{\tau}+\varepsilon;\ y,\bar{\tau})\right]\ \psi(y,\bar{\tau})\qquad (2)$$

is obtained from (1) [1-3]; where $\tilde{m}(y)$ is a suitable measure; $\hat{S}(x,\bar{\tau}+\varepsilon;\ y,\bar{\tau})$ is the action along the extremal joining $[y,\bar{\tau}]$ to $[x,\bar{\tau}+\varepsilon]$, for a fixed $x$ as $y$ varies over the manifold $\mathcal{R}_N$, with $\bar{\tau}=\tau(y)$ and $(\bar{\tau}+\varepsilon)=\tau(x)$. The corresponding differential equation for $\psi$ is obtained by expanding both sides of (2) in powers of $\sqrt{\varepsilon}$ and comparing the first few terms. If $d\tilde{m}(y)/dy$ depends on $y$, then it is often more convenient to obtain an equation for $\psi'$, where $\psi'(x)=(d\tilde{m}(x)/dx)\psi(x)$, and deduce equation for $\psi(x)$ by substitution.

Other alternative formulations of the path integral method are available in literature, e.g., for the propagator. However, all forms are essentially equivalent, and they are deducible from each other. Following analysis is applicable and the same results are deducible with all such forms, but the present one, representation of the wavefunction, is more suitable for the deductions to follow, which is also the original form.

### 1.2. Path Integral Formulation in Weyl's Geometry

Slightly adjusted form of (1) was obtained recently by a formulation on the background of Weyl's geometry [7, arXiv:1405.7693v2]. Weyl [8] assumed that the length of a vector $l_x$ changes as it is parallel transported from a point $y$ to $x$ along a trajectory $\rho_{xy}$. In addition, the length can be recalibrated at every point by multiplying $l_x$ at each $x$ by a function $\kappa(x)$, i.e., $l_x \to \kappa(x)l_x$. The net result is that a vector $l_y$ at $y$ transforms to $l_x$ at $x$, given by

$$l_x\ =\ \kappa(x)\ \exp\left[a\int_{\rho_{xy}} \phi_\mu dx^\mu\right]\kappa^{-1}(y)\ l_y,\qquad (3)$$

where $a$ is a nonzero constant; and $\phi_\mu$, termed the Weyl gauge potentials, are the components of a vector. The transformation $l_y \to l_x$ defined by (3), called gauge transformation, contains a point function $\kappa(x)$ of $x$ referred to as the assigned gauge, and the remaining exponential, path-dependent part, a functional, termed the essential gauge. The gauge transformations constitute the foundation of Weyl's geometry.

Let $\phi_\mu = aS_{,\mu}$, where $S(\rho_{xy})$ is the classical action along $\rho_{xy}$, and $,\mu$ denotes the derivative with respect to $x^\mu$. Physical trajectories facilitating the passage of a particle are defined by

$$\kappa(x) \exp\left[a\, S(\rho_{xy})\right] \kappa^{-1}(y) = 1, \qquad (4)$$

equivalently, $l_x = l_y$. Without loss of generality, lengths can be measured in units of $l_y$; then (4) reduces to $l_x = 1$. It was shown in [7] that (4) yields $a = i$, and since a constant cancels out in (4), $|\kappa(x')| = 1$ for all $x'$. Assigned gauges for two physically equivalent points were assumed to be equal; otherwise not. Also, the assigned gauges were defined in terms of the configuration and interaction of the observing system with the observed one; and their computation was illustrated with examples. Elemental path is defined to be the minimal solution of (4). A particle travels along a randomly selected elemental, and then along another elemental starting at the terminal point of the first. For $\kappa = 1$, an elemental is defined by $S = \pm 2\pi$. All monotonic paths are constituted of continuing unions of the elementals, and nonmonotonic paths are continuing unions of the monotonic ones.

In view of (4), (1) was adjusted to

$$\psi[\kappa; x, \tau(x)] = \sum_{\text{all paths}} \kappa(x) \exp\{i\, S[\rho_{xx_0}(\tau)]\} \kappa^{-1}(x_0). \qquad (5)$$

The representation of (5) is valid with each $\rho_{xx_0}$ constituting a segment of a physical trajectory. However, all trajectories in a simply connected region in an analytic manifold satisfy this condition. The cases where this does not hold require some adjustment to the analysis. For now, we restrict to the cases where the condition is satisfied. Clearly, the sum in (5) reduces to that on the elemental paths containing $x$, which is also equal to the sum over the continuing unions of elementals with the last one containing $x$.

It follows from (3) that with $a\phi'_\mu = iS_{,\mu}$, each term in (5) is equal to the length at $[x, \tau(x)]$ of a unit vector transported from $[x_0, \tau(x_0)]$ along $\rho_{xx_0}(\tau)$. Thus, (5) defines the wavefunction as an aggregate of the Weyl lengths at $[x, \tau(x)]$ of a unit vector transported to this point along all physical trajectories from everywhere; the length is defined by (3) with $a\phi'_\mu = iS_{,\mu}$. This provides a geomerical interpretation of the wavefunction.

The consequent counterpart of (2) reads

$$\kappa^{-1}(x)\psi(\kappa; x, \bar{\tau} + \varepsilon) = \int d\tilde{m}(y) \exp\left[i\, \hat{S}(x, \bar{\tau} + \varepsilon;\ y, \bar{\tau})\right] \kappa^{-1}(y)\psi(\kappa; y, \bar{\tau}) \qquad (6)$$

Comparison of (5) and (6) with (1) and (2), respectively shows that Feynman's form corresponds to $\kappa = 1$; for $\kappa \neq 1$, $\kappa^{-1}\psi(\kappa) = \psi$, and the corresponding results can be obtained from Feynman's formulation by substituting $\kappa^{-1}\psi(\kappa)$ for $\psi$. With this understanding, we restrict the following considerations to Feynman's representations (1) and (2) considered as special cases of (5) and (6), respectively.

**1.3. Klein-Gordon Equation**

Classical extremal, a geodesic, for the motion of a particle can be obtained by its variational characterization with the Lagrangian $m\sqrt{\dot{x}^\mu \dot{x}_\mu}$, where the dot denotes the derivative with respect to the parameter used to parametrize the curves. Arclength of the curve itself is usually used to parametrize it to determine the geodesic and thus, a geodesic is parametrized by its own arclength. Also, to determine a geodesic, $m$ can be taken to be an arbitrary nonzero constant but the action depends on it as it is obtained by integrating $m\sqrt{\dot{x}^\mu \dot{x}_\mu}$ along the curve. For now, we take $m$ to be the mass of particle but on occasions, some other parameter characterizing the particle will replace $m$ to obtain parallel results.

Due to unsuitability of a homogeneous Lagrangian for an application of the methods of path integration, a compatible inhomogeneous Lagrangian is used to calculate the action $\hat{S}$ along the extremal needed in (2). The action for the corresponding homogeneous Lagrangian can be obtained by requiring $\partial \hat{S} / \partial \varepsilon = 0$ to eliminate the parameter. The Lagrangian

$$L(x, \dot{x}) = \frac{m}{2}\left(\dot{x}^\mu \dot{x}_\mu + 1\right) = \frac{m}{2}\left(g_{\mu\nu}\dot{x}^\mu \dot{x}^\nu + 1\right) \tag{7}$$

satisfies the required conditions [6].

This Lagrangian yields a Schrödinger type differential equation for $\psi$ [6. arXiv1405.7693v1]:

$$i\frac{\partial \psi}{\partial \tau} = -\frac{1}{2m}\left[\partial_\mu \partial^\mu + m^2 - \frac{1}{3}R\right]\psi, \tag{8}$$

where $\partial_\mu$ denotes the covariant derivative in $\mathcal{R}_N$ and $R$ is its curvature scalar. The corresponding equation for the action calculated from the homogeneous Lagrangian $m\sqrt{\dot{x}^\mu \dot{x}_\mu}$, is obtained by setting $\partial \hat{S} / \partial \varepsilon = 0$, which implies that $\partial \psi / \partial \bar{\tau} = 0$. This reduces (8) to the Klein-Gordon equation generalized to the Riemannian spaces [9, arXiv 1405.7693v1]:

$$\left[\partial_\mu \partial^\mu + m^2 - R'\right]\psi(m) = 0, \tag{9}$$

where $R' = R/3$. In addition to providing the quantum mechanical description of the motion of a particle characterized by $m$, several physical systems can be described by the Klein-Gordon equation (9) as its special cases including a particle coupled to a general gauge field [10].

**1.4. Outline of Article**

In Sec. 2, a one parameter family of wavefunctions is generated from the translations of $\tau(x)$ in (1), which is shown to be periodic with respect to the translation parameter resulting from the periodicity of the phase-factors with respect to the action. Also, the quantized potentials with the particle being its fundamental quantum is constructed from the solutions of the Klein-Gordon equation, followed by deduction of the potential equation. In Sec. 3, an extended manifold endowed with a Riemannian structure $\mathcal{R}_{N+1}$ is constructed by adjoining the translation parameter to $\mathcal{R}_N$, and a quantized field formulation is developed in $\mathcal{R}_{N+1}$. In absence of the particles, the resulting total energy is equal to zero. Properties of the extended manifold are discussed in Sec. 4 yielding the classical description in $\mathcal{R}_{N+1}$ of essentially the same particle. This program can be carried out with $\mathcal{R}_{N+1}$ replacing $\mathcal{R}_N$ continuing this treatment to all $N$ generating the corresponding manifolds in the process.

In Sec. 5, the results are illustrated with about the simplest example, that of $\mathcal{R}_N$ being the 3D Euclidean space. It is shown that with the translation parameter identified with time, this yields the classical motion of a photon; and the corresponding Klein-Gordon equation provides its quantum mechanical description. The extended manifold is identified with the 4D Minkowski spacetime. The consequent field description in the 4D Minkowski spacetime results in the quantized Maxwell's equations with vacuum energy being equal to zero. In Sec. 6, the same program starting with the Minkowski manifold as the base is shown to yield parallel results for a massive particle defined by the photon. The paper is concluded with some concluding remarks in Sec. 7.

## 2. Translated Wavefunctions

Translate $x_0$ on each trajectory $\rho_{xx_0 x_0'}$ in (1) to $x_0'$ with arclength $\tau_{x_0 x_0'} = \tau'$, which results in the translation $\tau(x) \to \tau(x) + \tau'$ generating a one parameter family $\psi[x, \tau(x) + \tau']$ of the solutions of (1), equivalently of (2), i.e.,

$$\psi[x,\ \tau(x) + \tau'] \ =\ \sum_{\rho_{xx_0 x_0'}} \exp\{\ i\ S[\rho_{xx_0 x_0'}(\tau)]\}\ . \tag{10}$$

The arclength $\tau_{zy}$ of each trajectory $\rho_{zy}$ from a point $y$ to $z$ in $\mathcal{R}_N$ is given by

$$\tau_{zy}\ =\ \int_{\rho_{zy}} \sqrt{dx_\mu dx^\mu}\ =\ \int_{\rho_{zy}} d\tau\ \sqrt{\dot{x}_\mu \dot{x}^\mu}\ =\ S(\rho_{zy}; m)/m\ . \tag{11}$$

Since

$$S[\rho_{xx_0 x_0'}(\tau)]\ =\ \{S[\rho_{xx_0}(\tau)]\ +\ S[\rho_{x_0 x_0'}(\tau)]\}\ ,$$

it follows from (11) that the translations $\tau(x) \to [\tau(x) + 2n\pi/m]$ of arclength induce translations $S \to (S + 2n\pi)$ of the action, where $n$ is an arbitrary integer. Since (1) is invariant under the translations $S \to (S + 2n\pi)$, the contribution of each trajectory to the wavefunction is also invariant under these corresponding translations of the arclength. This is essentially a statement of the periodicity of phase-factors with respect to the action together with proportionality of the action and arclength given by (11). It follows that $\psi[x, \tau(x) + \tau']$ is a periodic function of $\tau'$ with its minimal period being equal to $2\pi/m$, which is equal to the arclength $\tau_m^0$ of the corresponding elemental trajectory $\rho_m^0$ defined by $S(\rho_m^0; m) = 2\pi$.

In general, a periodic function $\phi(x, \tau')$ of $\tau'$ with period equal to $2\pi/m$ admits a Fourier series expansion:

$$\phi(x, \tau') = \sum_{n=-\infty}^{\infty} \phi_n(x) \exp[inm\tau'], \qquad (12)$$

with $2\pi/(nm)$ being the minimal period for each term $\{\phi_n(x) \exp[inm\tau']\}$. Since the minimal period of $\psi[x, \tau(x) + \tau']$ is equal to $(2\pi/m)$, it is expressible as

$$\psi[x, \tau(x) + \tau'] = \psi[x, \tau(x)] \exp[im\tau'].$$

As discussed in Sec. 1.3, with the action in (1) and (2) computed from the Lagrangian $m\sqrt{\dot{x}^\mu \dot{x}_\mu}$, we have $\psi[x, \tau(x)] = \psi(m)$, where $\psi(m)$ is the solution of (9). This reduces $\psi[x, \tau(x) + \tau']$ further to $\{\psi(m) \exp[im\tau']\}$. Thus, with $\phi_1(x) = \psi(m)$, $\psi[x, \tau(x) + \tau']$ is the term for $n=1$ in (12). Having identified the term corresponding to $n=1$ with the family of the quantum mechanical wavefunctions generated by the translations of arclength for the motion of particle characterized by $m$, in the following we complete the series given by (12).

Since $S(\rho_m^0; m) = 2\pi$, it is clear from (11) that $S(\rho_m^0; nm) = nS(\rho_m^0; m) = 2n\pi$ for each integer $n \neq 0$ defining the minimal period $2\pi/(nm)$ corresponding to $nm$, which is the arclength of the elemental segment $\rho_{nm}^0$ of $\rho_m^0$ associated with the particle characterized by $nm$. It is clear that for $n \neq 0$, $\rho_m^0$ and $\{\rho_{nm}^0\}$ imply existence of each other as each $\rho_{nm}^0$ is an integral fractional segment of $\rho_m^0$ in the sense that $S(\rho_{nm}^0; m) = 2\pi/n$, and $\rho_m^0$ can be constructed as a continuing union of $n$ members of $\{\rho_{nm}^0\}$. Thus, (12) represents the quantized collection of particles characterized by $\{nm\}_{n=-\infty}^{\infty}$, termed the $m$-class.

For $n=0$, we have $S(\rho; nm) = S(\rho; 0) = 0$ for each trajectory $\rho$. It follows from this together with the definition $S(\rho_0^0; nm) = 2\pi$ of the elemental trajectory $\rho_0^0$ corresponding to $nm=0$, that no elemental trajectory, and hence no contributing trajectory, exits for $n=0$, implying that $\phi_0 = 0$.

The solutions $\psi(nm) = \phi_n(x)$ can be obtained by substitutions in (9), i.e.,

$$\left[\partial_\mu \partial^\mu + (nm)^2 - R'\right] \phi_n = 0. \tag{13}$$

Let

$$\hat{\phi}(x, \tau') = \sum_{n=-\infty}^{\infty} \hat{\phi}_n(x) \exp[inm\tau'], \tag{14}$$

where $\hat{\phi}_n$ are the solutions of

$$\left[\partial_\mu \partial^\mu + (nm)^2\right] \hat{\phi}_n = 0,$$

which can be obtained, e.g., by the perturbation method starting with the approximate solution $\hat{\phi}_n = \psi(nm)$. It is clear that $\hat{\phi}$ defined by (14) satisfies the equation

$$\left[\frac{\partial^2}{\partial \tau'^2} - \partial_\mu \partial^\mu\right] \hat{\phi} = 0. \tag{15}$$

Conversely, (15) together with the periodic boundary condition $\hat{\phi}(x, 0) = \hat{\phi}(x, 2\pi/m)$ yields (14), which in turn yields the collection $\{\hat{\phi}_n\}$, defining also $\{\phi_n\}$, describing the particles of $m$-class constituting the set of particles quantized in units of $m$. It follows that while $\phi_n$, equivalently $\hat{\phi}_n$, provide the quantum mechanical description of the particle characterized by $(nm)$ in the collection, $\hat{\phi}$ describes them collectively and thus, provides their field description. Clearly, the field is quantized with the particle characterized by $m$ being its fundamental quantum.

## 3. Fields

Let $\mathcal{R}_{N+1}$ be the extended manifold obtained by adjoining the $\tau'$-axis to $\mathcal{R}_N$ and endowing it with the metric $\hat{g}$ defined by

$$\hat{g}_{00} = 1;\ \hat{g}_{0\mu} = \hat{g}_{\mu 0} = 0;\ \hat{g}_{\mu\nu} = \hat{g}_{\nu\mu} = -g_{\mu\nu};\ \mu, \nu = 1, 2, ..., N, \tag{16}$$

which determines the infinitesimal arclength $d\hat{\tau}$ in $\mathcal{R}_{N+1}$ by

$$d\hat{\tau}^2 = \hat{g}_{\mu\nu} d\hat{x}^{\hat{\mu}} d\hat{x}^{\hat{\nu}} = d\hat{x}_\mu d\hat{x}^\mu,\ \hat{\mu}, \hat{\nu} = 0, 1, 2, ..., N,$$

with $\hat{x}^0 = \tau'$, $\hat{x}^\mu = x^\mu$, $\mu = 1, 2, ..., N$.

The potential equation (15) assumes a covariant form in $\mathcal{R}_{N+1}$:

$$\partial_{\hat{\mu}} \partial^{\hat{\mu}} \hat{\phi} = 0; \tag{17}$$

where $\partial_{\hat{\mu}}$ denotes the covariant derivative in $\mathcal{R}_{N+1}$.

The solution $\hat{\phi}$ of (17) obtained in Sec. 2 is a scalar function. There exist $(N+1)$ linearly independent scalar solutions of (17), e.g., $\{\phi'(\hat{\mu})\}$, where $\phi'(\hat{\mu})$ is a $\hat{x}^\mu$-independent solution of (17), i.e., with $\partial_{\hat{\mu}}\phi'(\hat{\mu}) = 0$ for each $\hat{\mu}$. Each solution of (17) is given by some linear combination of the elements of $\{\phi'(\hat{\mu})\}$. If desired, the elements of $\{\phi'(\hat{\mu})\}$ can be combined to obtain another linearly independent set of solutions, still denoted by $\{\phi'(\hat{\mu})\}$. This set can be used to define $(N+1)$-vector solutions of (17), as follows. Define $(N+1)$ vectors $\phi''^{\hat{\mu}}$ with its $\hat{\mu}$-th component being $\phi'(\hat{\mu})$ and all the others being equal to zero. It is clear that $\{\phi''(\hat{\mu})\}$ is a set of $(N+1)$ linearly independent vector solutions of (17). Yet again, each vector solution of (17) can be obtained as a linear combination of the elements of $\{\phi''(\hat{\mu})\}$. Freedom still left is removed by some additional conditions. In the following we assume that a suitable vector solution $\hat{\phi}$ of (17) is available, obtained by this or some other method. It can be seen that each component $\hat{\phi}^{\hat{\nu}}$ satisfies the potential equation (17).

Tensor fields are constructed from the derivatives of the vector $\hat{\phi}$. An antisymmetric tensor of second rank $f^{\mu\nu}$ in $\mathcal{R}_{N+1}$ and its dual $\hat{f}^{\hat{\mu}_0...\hat{\mu}_{N-2}}$ of rank $(N-1)$ are given by

$$f^{\hat{\mu}\hat{\nu}} = \left(\partial^{\hat{\nu}}\hat{\phi}^{\hat{\mu}} - \partial^{\hat{\mu}}\hat{\phi}^{\hat{\nu}}\right), \quad \hat{f}^{\hat{\mu}_0...\hat{\mu}_{N-2}} = \varepsilon^{\hat{\mu}_0...\hat{\mu}_N} f_{\hat{\mu}_{N-1}\hat{\mu}_N} \tag{18}$$

for all $\hat{\mu}$, $\hat{\nu}$, $\hat{\mu}_j$, $j = 0, 1, 2, ..., N$. Here $\varepsilon^{\hat{\mu}_0...\hat{\mu}_N}$ is the Levi-Civita tensor density.

By virtue of its definition, the dual satisfies the Jacobi identities:

$$\partial_{\hat{\mu}_j} \hat{f}^{\hat{\mu}_0...\hat{\mu}_j...\hat{\mu}_{N-2}} = 0, \tag{19}$$

which can be seen by permuting the superscripts in the Levi-Civita tensor density with substitutions from (18). The identity

$$\partial_{\hat{\nu}} f^{\hat{\mu}\hat{\nu}} = \partial_{\hat{\nu}} \left(\partial^{\hat{\nu}}\hat{\phi}^{\hat{\mu}} - \partial^{\hat{\mu}}\hat{\phi}^{\hat{\nu}}\right)$$

together with the extended Lorentz gauge fixing condition $\partial_{\hat{\mu}}\hat{\phi}^{\hat{\mu}} = 0$ and (17) for $\hat{\phi}^{\hat{\mu}}$ yields

$$\partial_{\hat{\nu}} f^{\hat{\mu}\hat{\nu}} = 0. \tag{20}$$

Conversely, (19) implies that $f^{\hat{\mu}\hat{\nu}}$ can be expressed as an antisymmetric derivative of a vector potential $\hat{\phi}$ as in (18), which together with the extended Lorentz gauge fixing condition yields the potential equation (17) for each $\hat{\phi}^{\hat{\mu}}$.

It should be noticed that (19) and (20) are homogeneous free field equations in the sense that there are no explicit sources and sinks although some interactions are implicitly included in them through the Riemannian structure of space. For example, the gauge fields are expressible in terms of the Riemannian space structures in the Kaluza-Klein formulation [10]. Classical trajectory of a particle coupled to a gauge field results as a geodesic in the corresponding Riemannian Kaluza-Klein space. Quantum description of such a particle is provided by the Klein-Gordon equation in the same space and its field description results from the above formulation. Similar arguments apply to the gravitational field. Also, a parallel formulation can be developed in terms of $\{\phi_n\}$ instead of $\{\hat{\phi}_n\}$, in which case a source term defined in terms of the curvature scalar would show up explicitly as well, in addition to the effects still included in the other terms. Higher rank tensor fields and the related field equations can be generated in a similar manner.

## 4. Extended Manifolds

With reference to the collection of curves $\{\rho_{yx}, \rho_{xy}\}$, i.e., with one of its end points at $x$, let the $\tau'$-axis be attached to $\mathcal{R}_N$ with its origin at $x$, i.e., the origin of $\mathcal{R}_{N+1}$ is defined to be at $(0, x)$. Such precision was not necessary for the purpose of Sec. 3; in this section, it will be seen to add substantial simplicity. For clarity, consider the trajectories $\{\rho_{yx}\}$, i.e., with initial point at the origin. Parallel results for $\{\rho_{xy}\}$ can be inferred or obtained similarly.

For a geodesic $\rho_{yx}^g$ in $\mathcal{R}_N$, which together with its arclength $\tau_{yx}^g$ is determined by its end points, (11) yields

$$\tau_{yx}^g = \int_{\rho_{yx}^g} \sqrt{dx_\mu dx^\mu} = \int_{\rho_{yx}^g} d\tau \sqrt{\dot{x}_\mu \dot{x}^\mu} = \frac{\hat{S}(y, x)}{m}, \tag{21}$$

where $\hat{S}(y, x)$ is the action along $\rho_{yx}^g$. The relation $\tau_{yx} = S(\rho_{yx})/m$ holds for each $\rho_{yx}$ but for non-geodesics, the value of $S(\rho_{yx})$ and the arclength $\tau_{yx}$ depend on the trajectory while for a geodesic, they are reducible to functions of the end points of $\rho_{yx}^g$. Incidentally, this property underlies the method described in Sec. 1.3 to evaluate $\hat{S}(y, x)$, which is by solving the Hamilton-Jacobi equation obtained from the Lagrangian defined by (7) and eliminating $\tau_{yx}$ by requiring $\partial \hat{S}(y, x; \tau_{yx})/\partial \tau_{yx} = 0$, where $\hat{S}$ is the action along the corresponding extremal.

The curve $\rho_{yx}^g$ is lifted to $\hat{\rho}_{yx}^g$ in $\mathcal{R}_{N+1}$ defined by the lift of each point $z$ on $\rho_{yx}^g$ to $(\tau_{zx}^g, z)$. This can be done for each point on each geodesic and thus, $d\hat{\tau}^2 = 0$ everywhere on each $\hat{\rho}_{yx}^g$, where $d\hat{\tau}^2$ is the infinitesimal arclength in $\mathcal{R}_{N+1}$ determined by (16). Clearly, (21) constitutes a restrictive equation between the coordinates $(\tau_{zx}^g, z)$ of each point of $\hat{\rho}_{yx}^g$ by expressing one as a function of the others. Thus, the collection of geodesics in $\mathcal{R}_N$ defines an $N$-dimensional surface $\mathcal{R}_{N+1}^{0+}$ in $\mathcal{R}_{N+1}$ accommodating the lifts $\hat{\rho}_{yx}^g$ of $\rho_{yx}^g$ with $\tau_{yx}^g \geq 0$. Its counterpart $\mathcal{R}_{N+1}^{0-}$ corresponding to the curves converging onto $x$, i.e., with $\tau_{yx}^g \leq 0$, is constructed similarly. The null surface $\mathcal{R}_{N+1}^{0}$ is the union of two surfaces $\mathcal{R}_{N+1}^{0+}$ and $\mathcal{R}_{N+1}^{0-}$ with common apex at $(0, x)$.

Above construction renders $\tau'$ to be a natural parameter to parametrize the curves in $\mathcal{R}_N$, but $\tau'$ is defined by the translations of $\tau$ used to parametrize these trajectories, i.e., $\tau'$ describes the corresponding translation group. However, $\tau'$ can be identified with $\tau$ for the purpose of the pertaining calculations, which is legitimate for $\tau'$ covers all arbitrary translations of arclength of each curve and thus, it can be defined in one to one correspondence with $\tau$, although $\tau'$ conceptually remains a group parameter. This construction is consistent with the usual parametrization of a geodesic by its own arclength as discussed in Sec. 1.3.

With identification of $\tau'$ with $\tau$, the velocities $\dot{x}_\mu$ of a particle moving along a geodesic, characterized by $d\hat{\tau}^2 = 0$ satisfy the relation $\dot{x}_\mu \dot{x}^\mu = 1$ describing a particle traveling with unit speed, and the consequent momenta $p_\mu = m\dot{x}_\mu$ are related by $p_\mu p^\mu = m^2$. For the other trajectories in $\mathcal{R}_{N+1}$, characterized by $d\hat{\tau}^2 \neq 0$, $\hat{\tau}$ is a more suitable parameter, yielding $\sqrt{1 - \dot{x}_\mu \dot{x}^\mu} \neq 0$. Since $d\hat{\tau} = d\tau' \sqrt{1 - \dot{x}_\mu \dot{x}^\mu}$, the associated velocities $x_\mu^\bullet$ are given by $x_\mu^\bullet = \dot{x}_\mu / \sqrt{1 - \dot{x}_\mu \dot{x}^\mu}$, where the thick dot superscript denotes the derivative with respect to $\hat{\tau}$. The momenta $\hat{p}_\mu = m x_\mu^\bullet$ in $\mathcal{R}_{N+1}$ are related to $p_\mu$ by

$$\hat{p}_\mu = m\dot{x}_\mu / \sqrt{1 - \dot{x}_\mu \dot{x}^\mu} = p_\mu / \sqrt{1 - \dot{x}_\mu \dot{x}^\mu}, \qquad (22)$$

naturally generating an additional momentum $\hat{p}_0 = m / \sqrt{1 - \dot{x}_\mu \dot{x}^\mu}$.

Starting with the Lagrangian classically describing a particle characterized by an arbitrary constant $m$ in $\mathcal{R}_N$, we developed its quantum mechanical particle formulation in $\mathcal{R}_N$ in the framework of the path integral formulation. Then, the family of wavefunctions generated by the translations of arclength was used to deduce the potential equation, which was expressed in $\mathcal{R}_{N+1}$ constructed by augmenting $\mathcal{R}_N$ by attaching the translation parameter to it together with a suitable metric. The field tensors and the corresponding equations are deduced from the potentials and the equation they satisfy, yielding the quantized field description in $\mathcal{R}_{N+1}$ with the same particle as the quantum of the field. This program can be repeated by taking $\mathcal{R}_{N+1}$ for $\mathcal{R}_N$ and thus, formulate the classical motion in $\mathcal{R}_{N+1}$ of a particle still characterized by the parameter $m$, based on the Lagrangian $m\sqrt{\dot{x}_{\mu}\dot{x}^{\mu}}$ and its inhomogeneous counterpart defined by (7). Results of Sec. 1.3 provide the quantum mechanical particle description in $\mathcal{R}_{N+1}$. Formulations for potentials and fields in $\mathcal{R}_{N+2}$, generated in the process, are obtained by the methods of Sec. 3.

Clearly, this program continues ad infinitum.

## 5. Electromagnetic Field

In this section, we illustrate the results of earlier sections by considering about the simplest example, that of a particle characterized by $m = \omega$ in 3D Euclidean space $\mathcal{E}_3$ which is a flat manifold with its metric being the identity matrix. The results can be obtained by substituting $\mathcal{E}_3$ for $\mathcal{R}_N$. Classically, the particle can be described by the Lagrangian $\omega\sqrt{\dot{x}_{\mu}\dot{x}^{\mu}}$, $\mu = 1, 2, 3$, with arclength as the parameter. As discussed in Sec. 1.3, the Lagrangian given by (7) is more suitable for an application of the path integral methods. The trajectories in $\mathcal{E}_3$ are parametrized by the arclength measured from a reference point. In Sec. 2, the translation group parameter is defined by the translations of the arclength, which, from Sec. 4, can be identified with the arclength for the classical treatment. With the arclength of the curve as the parameter, the Lagrangian $\omega\sqrt{\dot{x}_{\mu}\dot{x}^{\mu}}$ describes a particle travelling in a straight line in $\mathcal{E}_3$. Identifying the translation parameter with time $t$, we have $\dot{x}_{\mu}\dot{x}^{\mu} = 1$, which describes a particle traveling with unit speed.

From the definition of the Lagrangian, the momenta are given by

$$p_{\mu} = \omega\dot{x}_{\mu} = \omega\dot{x}^{\mu} = p^{\mu}$$

yielding the relation $p_{\mu}p^{\mu} = \omega^2$, which is identified with the energy-momentum relation for the photon of energy $\omega$ in natural units with the speed of light being taken to be equal to one. This together with the above result that the particle travels along a straight line with unit speed, provides the classical particle description of the photon of energy $\omega$. With proper substitutions, the Klein-Gordon equation (9),

$$\left[\partial_{\mu}\partial^{\mu} + \omega^2\right]\psi(\omega) = \left[\nabla^2 + \omega^2\right]\psi(\omega) = 0, \tag{23}$$

provides the quantum particle description of the same photon.

The discussion following (12) yields the photons with energies of $(n\omega)$ for all nonzero integers $n$. With proper substitutions, the $n$-photon solutions are given by (13), i.e.,

$$\left[\nabla^2 + (n\omega)^2\right]\phi_n = 0, \quad n = \pm 1, \pm 2,\ldots. \tag{24}$$

For $n=0$, i.e., no photon, $n\omega=0$, and based on the fact that there are no contributing trajectories for $n=0$, $\phi_0=0$, i.e., there is no field, as discussed in Sec. 2. Thus, vacuum has zero energy. This follows from (24) also as $\nabla^2\phi_0=0$ implies that $\phi_0=0$.

The extended manifold is obtained by adjoining the translation parameter identified with time $t$ to $E_3$ as the additional coordinate $\hat{x}^0$. With the metric defined by (16), this yields the 4D Minkowski spacetime $M_4$. Globally, the photon path is given by $t^2 - x_\mu x^\mu = 0$, corresponding to the straight lines in $E_3$. The null manifold $M_4^0$ yielded by the construction of Sec. (4), is the 3D surface of the light cone.

From (14), the potentials $\phi = \hat{\phi}$ are defined by

$$\phi(x,t) = \sum_{n=-\infty}^{\infty} \phi_n(x) \exp(in\omega t). \tag{25}$$

From (15) and (17), the 4-vector potential $\phi$ is a solution of the equation

$$\left[\frac{\partial^2}{\partial t^2} - \nabla^2\right]\phi = \partial_{\hat{\mu}}\partial^{\hat{\mu}}\phi = 0, \quad \hat{\mu} = 0,1,2,3, \tag{26}$$

which is identified with the equation for the electromagnetic potentials with known solutions.

Tensor fields of second rank can now be constructed by differentiations of the vector $\phi = \hat{\phi}$. From (18), the antisymmetric tensor $f^{\hat{\mu}\hat{\nu}}$ and its dual $\hat{f}^{\hat{\mu}\hat{\nu}}$, both of second rank in $M_4$, are given by

$$f^{\hat{\mu}\hat{\nu}} = \left(\partial^{\hat{\nu}}\hat{\phi}^{\hat{\mu}} - \partial^{\hat{\mu}}\hat{\phi}^{\hat{\nu}}\right), \quad \hat{f}^{\hat{\mu}\hat{\nu}} = \varepsilon^{\hat{\mu}\hat{\nu}\hat{\sigma}\hat{\eta}} f_{\hat{\sigma}\hat{\eta}}; \quad \hat{\mu},\hat{\nu},\hat{\sigma},\hat{\eta} = 0,1,2,3. \tag{27}$$

As a special case of (19), the dual satisfies the Jacobi identities:

$$\partial_{\hat{\nu}}\hat{f}^{\hat{\mu}\hat{\nu}} = 0, \tag{28}$$

which constitute one pair of the Maxwell equations. The other pair of the free field Maxwell's equations

$$\partial_{\hat{\nu}} f^{\hat{\mu}\hat{\nu}} = 0, \tag{29}$$

follows from (20). It is clear from the derivations that the potentials and fields in the above equations are quantized.

Above treatment of electromagnetism follows the reverse procedure of the standard one, which starts with a pair of classical Maxwell's equations formally the same as (28) that imply the existence of a vector potential related to the fields by (27) satisfying (26). Space Fourier transform of (26) yields the equation for a classical simple harmonic oscillator of unit mass. On this basis, the electromagnetic field is considered a collection of infinitely many simple harmonic oscillators. Second quantization replaces the classical oscillators with the quantum ones. Since the ground state energy of a quantum simple harmonic oscillator is strictly positive, this procedure yields infinite vacuum energy. Differences in the energy levels still remain finite as the vacuum energy cancels out. Arguing that the experiments determine only the differences in energy levels, not their absolute values, and the differences so calculated agree with the observations, this formulation is accepted. However, this does not eliminate the problem resulting from infinite vacuum energy as discussed below [1, pp. 244-246].

Energy is equivalent to mass, which generates a proportionate gravitational field and thus, infinite vacuum energy generates an unphysical infinite gravitational field. Due to some observational disagreements with electrodynamics at high energies, revealing the shortcomings of the existing formulation, a questionable large energy cut off is used hoping to correct the problem of infinite gravitational field together with some others. However, the predicted gravitational field is still too large to escape detection, which is not observed. The present formulation yields zero for the vacuum energy while retaining the satisfactory results of the prevailing theory, i.e., the energy differences.

Some phenomena are described by accepting infinite or a large vacuum energy of the existing formulation. However, one contradiction is sufficient to invalidate a theory even if it may describe the other observations satisfactorily. The present formulation eliminates the problem of large vacuum energy eliminating the associated contradiction without impacting upon the energy differences. What is required now is to describe the other phenomena also within the framework of the present formulation, which is outside the scope of the present paper where the basic formulation is presented, although desirable for future studies.

## 6. Field of Massive Particle

In Sec. 5, we developed the classical and quantum mechanical particle formulations and the field formulation of a photon. In this section, we describe parallel formulation of the same particle in $\mathcal{M}_4$.

From (22), the corresponding momenta $\hat{p}_{\hat{\mu}}$ in $\mathcal{M}_4$ are given by $\omega \dot{x}_\mu / \sqrt{1 - \dot{x}_\mu \dot{x}^\mu}$, $\hat{p}_0 = \omega / \sqrt{1 - \dot{x}_\mu \dot{x}^\mu}$ yielding $\hat{p}_{\hat{\mu}} \hat{p}^{\hat{\mu}} = \omega^2$, which is the relativistic energy-momentum relation for a particle of rest mass $m = \omega$, with the fourth momentum $\hat{p}_0$ being its total energy.

Counterpart of the Lagrangian $\omega \sqrt{\dot{x}_\mu \dot{x}^\mu}$ in $\mathcal{M}_4$ is clearly $m \sqrt{\dot{\hat{x}}_{\hat{\mu}} \dot{\hat{x}}^{\hat{\mu}}}$ providing the classical description of the motion of particle. For convenience, we adjust the notation to express this Lagrangian in $\mathcal{M}_4$ as $m \sqrt{\dot{x}_\mu \dot{x}^\mu}$, $\mu = 1, 2, 3, 4$, as well as adjust the notation elsewhere accordingly. The arclength in $\mathcal{M}_4$ and the translation parameter are denoted by $\tau$ and $\tau'$, respectively. Quantum mechanical description of this particle of mass $m$ is given by (9) and for the collection of particles in the $m$-class, by (13), i.e.,

$$\left[\partial_\mu \partial^\mu + (nm)^2\right] \hat{\phi}_n = 0, \quad n = \pm 1, \pm 2, \ldots \tag{30}$$

where $\hat{\phi}_n = \psi(nm)$. From (14), the potential is expressed as

$$\hat{\phi}(x, \tau') = \sum_{n=-\infty}^{\infty} \hat{\phi}_n(x) \exp(inm\tau').$$

From (15) and (17) the 5-vector potential $\hat{\phi}$ is obtained by solving

$$\left[\frac{\partial^2}{\partial \tau'^2} - \partial_\mu \partial^\mu\right] \hat{\phi} = \partial_{\hat{\mu}} \partial^{\hat{\mu}} \hat{\phi} = 0, \tag{31}$$

with $\hat{\mu} = 0, 1, 2, 3, 4$ and $\hat{x}_0 = \hat{x}^0 = \tau'$. From (18), the field tensors are given by

$$f^{\hat{\mu}\hat{\nu}} = \left(\partial^{\hat{\nu}} \hat{\phi}^{\hat{\mu}} - \partial^{\hat{\mu}} \hat{\phi}^{\hat{\nu}}\right); \quad \hat{f}^{\hat{\mu}_0 \hat{\mu}_1 \hat{\mu}_2} = \varepsilon^{\hat{\mu}_0, \hat{\mu}_1, \hat{\mu}_2, \hat{\mu}_3, \hat{\mu}_4} f_{\hat{\mu}_3 \hat{\mu}_4};$$
$$\hat{\mu}, \hat{\nu}, \hat{\mu}_j = 0, 1, 2, 3, 4; \quad j = 0, 1, 2, 3, 4. \tag{32}$$

The field equations follow from (19) and (20) to be

$$\partial_{\hat{\mu}_j} \hat{f}^{\hat{\mu}_0 \hat{\mu}_1 \hat{\mu}_2} = 0; \quad \partial_{\hat{\nu}} f^{\hat{\mu}\hat{\nu}} = 0. \tag{33}$$

This process continues for higher dimensions, step by step.

## 7. Concluding Remarks

Classical action with Hamilton's action principle and path integral formulation, respectively, determines the classical and quantum behavior of a particle. In the present article, a family of wavefunctions is generated by the translations of arclength used to parameterize the paths; and used to formulate the quantized fields with the particle being its basic quantum. Thus, the present treatment formulates the classical and quantum attributes of a particle, as well as the associated field description in a unified framework. The field in this formulation has zero vacuum energy eliminating the problem of an unphysical infinity. This procedure can be repeated with the extended manifold replacing the original one to continue this program step by step in higher dimensional spaces generated in the process.

Results are illustrated with base being the 3D Euclidean space yielding the classical and quantum particle descriptions of a photon in the base space, and its field description in the 4D Minkowski spacetime generated in the process. The field formulation results in the quantized Maxwell equations. A novel interpretation of time results in the process. Illustration is continued with repeating this program in the Minkowski spacetime for a massive particle defined by the photon. Classical description of this particle in the present formulation is the same as its relativistic description. This brings the special relativity also in the same unified framework. In fact, a generalized form of the relativistic formulation results from the classical description in the general extended manifold. Results parallel to the photon are obtained also for this massive particle.

Present formulation motivates and provides the foundation for further studies. We mention a few obvious ones: Application of the general formulation of fields to 3D Euclidean space without charges yields the corresponding quantized Maxwell's equations. This formulation can be employed to deduce the quantized field equations in other Riemannian spaces, e.g., the spaces in the presence of the gauge fields, and gravitation. Present construction generates higher dimensional manifolds and particles together with the corresponding field equations. Physical significance of the higher dimensional manifolds together with the related particles warrants further studies.